\documentclass[aps,prl,showpacs,twocolumn]{revtex4}
\usepackage{graphicx}
\usepackage{amsfonts}
\usepackage{xcolor}
\usepackage[normalem]{ulem}
\usepackage{bm}

\begin{document}

\title{Chern Insulators on Singular Geometries}
\author{Ai-Lei He$^{1}$, Wei-Wei Luo$^{1}$, Yi-Fei Wang$^{2}$, Chang-De Gong$^{2,1}$ }
\affiliation{$^1$National Laboratory of Solid State Microstructures and Department of Physics, Nanjing University, Nanjing 210093, China  \\$^2$Center for Statistical and Theoretical Condensed Matter Physics, and Department of Physics, Zhejiang Normal University, Jinhua 321004, China}
\date{\today}

\begin{abstract}
Topological quantum states have been proposed and investigated on two-dimensional flat surfaces or lattices with different geometries like the plane, cylinder and torus. Here, we study quantum anomalous Hall (QAH) or Chern insulator (CI) states on two-dimensional singular surfaces (such as conical and helicoid-like surfaces). Such singular geometries can be constructed based on the disk geometry and a defined unit sector with $n$-fold rotational symmetry. The singular geometry induces novel and intriguing features of CI/QAH states, such as in-gap and in-band core states, charge fractionalization, and multiple branches of edge excitations.
\end{abstract}


\maketitle

{\it Introduction.---}The discovery of integer quantum Hall (IQH) states~\cite{Klitzing} with integer Hall conductances is a miracle in condensed matter physics which introduces the notion of topology into physics~\cite{Thouless}. The Haldane model~\cite{Haldane} is the first quantum anomalous Hall (QAH) or Chern insulator (CI) model with non-trivial topological bands labeled by Chern numbers~\cite{Thouless}. Other CI/QAH models were proposed subsequently, like the checkerboard-lattice model, the kagom{\'e}-lattice model, and the lattice Dirac model~\cite{Yakovenko,Nagaosa,Lattice,Kagome,Checkerboard}. Recently, the Haldane model has been realized in ultracold fermionic systems~\cite{RealHaldane}. The IQH state can be described based on the Pauli principle with an integer filling factor and its wave function is a single Slater determinant. Clearly, this is much simpler than the fractional quantum Hall (FQH) states~\cite{Tsui,Laughlin} which obey the generalized Pauli principle (GPP)~\cite{GPP}.

\begin{figure}[!htb]
  \vspace{-0.05in}
\includegraphics[scale=0.45]{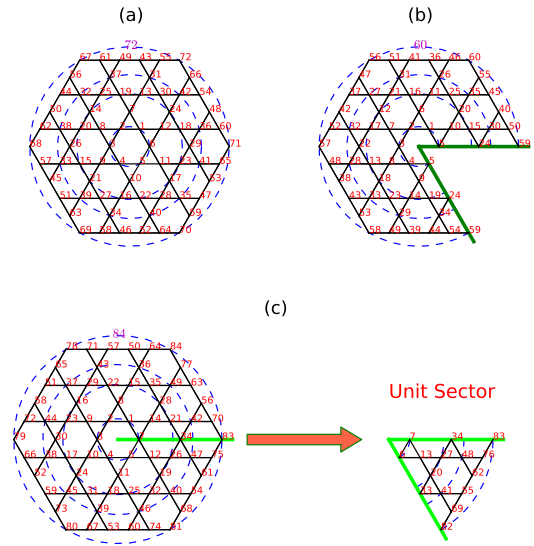}
  \vspace{-0.2in}
\caption{(color online).  The kagom\'{e} lattice on planar and singular surfaces. (a) On disk geometry, the kagom\'{e} lattice satisfies the $C_6$ rotational symmetry. (b) The unfolded drawing of the $C_5$ rotational symmetric kagom\'{e} lattice on the conical surface. Once reducing the rotational symmetry of the disk, the lattice is on the conical surface. (c) The unfolded drawing of the $C_7$ rotational symmetric kagom\'{e} lattice is viewed as a disk geometry gluing a unit sector. The lattice size is indicated by the circles and the labelled numbers. We mark the lattice sites with 1,2,3... to highlight the connections.}
\label{disk_Lattice}
\end{figure}

\begin{figure*}
  \vspace{-0.1in}
\includegraphics[width=14cm]{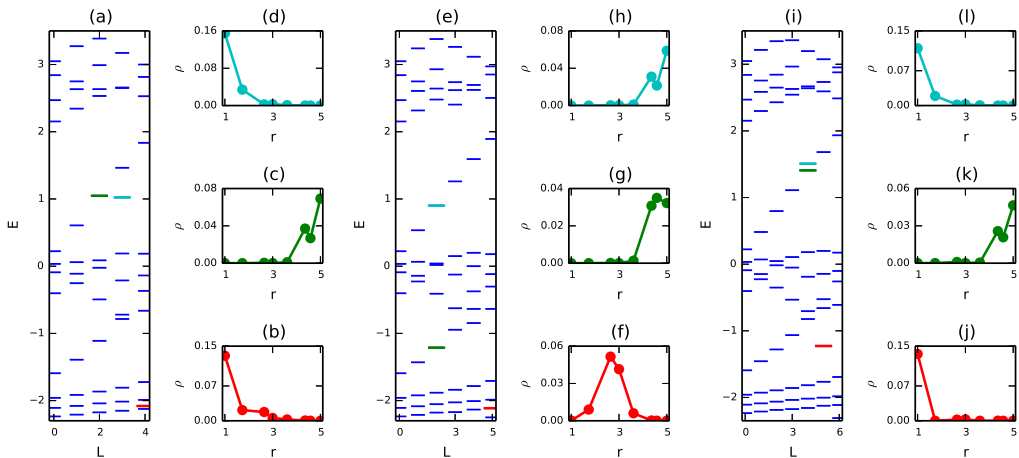}
  \vspace{-0.2in}
\caption{(color online). (a)-(d) The energy and density profile of the $60$-site kagom{\'e} lattice with $5$-fold rotational symmetry on the conical surface. The core states are marked by the red and cyan short line. The density of these two states is shown in (b) and (d) with a local density near the corner and the edge state is show in (c) with the green line. (e)-(h) The energy and density profile of the $72$-site kagom{\'e} lattice with $6$-fold rotational symmetry on disk. We show the density of bulk and edge states in (f)-(h). For the disk, there is no core states in the gap. (i)-(l) The energy and density profile of $84$-site kagom{\'e} lattice with $7$-fold rotational symmetry on the singular surface. (j)-(l) The localized in-gap states. (k) The edge state. Here, we choose a trap potential $V_{\rm trap}=0.005$.}
\label{single_particle_states}
\end{figure*}

Charge fractionalization is one of the key features of FQH states, however, fractional charge can also be observed in IQH systems~\cite{Milletari,Inoue}. Fractional charge is also predicted in some low-dimensional systems like the Su-Schrieffer-Hegger (SSH)~\cite{SSH,SSH1} model, the Kekul{\'e} graphene structures~\cite{Hou}, the IQH states on the conical surface~\cite{Biswasa,Klevtsov1} and defects in topological lattices~\cite{Ruegg,Moore}. Fascinatingly, the FQH states not only can be investigated on various geometries such as the disk, sphere, cylinder and torus~\cite{Laughlin,FQH_Sphere,FQH_Cylinder,FQH_Torus}, but also can be constructed on the singular surfaces~\cite{TuHH,Klevtsov2,TCan1,TCan2,Gromv}. For FQH states, the excessive charges on the tip of the conical surface can be estimated~\cite{TuHH}. Similarly,  the CI/QAH states, as well as the fractional quantum anomalous Hall (FQAH) or fractional Chern insulator (FCI) states~\cite{FCI_Geometry,FCI_Geometry1,WWLuo,ALHe} have been studied on the disk, cylinder and torus. Recently, topological states were proposed on the singular surfaces such as the fullerenes~\cite{Moore} and M\"{o}bius surfaces~\cite{Mobius1,Mobius2}. The bulk-edge correspondence is another key feature of IQH/FQH states. Edge excitations of IQH states were initially described by the 1D Fermi liquid~\cite{Halperin}, while edge excitations for FQH states are described by the 1D chiral Luttinger liquid~\cite{Wen1,Wen2}. There is one branch of edge excitations in the $\nu=1$ and $\nu=1/m$ quantum Hall states and two branches of edge excitations in the FQH states with filling factor $1-1/m$ and $2/5$ ($m$ is an odd integer)~\cite{Wen1}. Edge excitations for FCI/FQAH states have been previously obtained and reconstructed based on the GPP~\cite{WWLuo,ALHe}.

In this paper, we construct the CI/QAH states on singular lattices with arbitrary $n$-fold rotational symmetry. There are some localized states around the lattice center (core states) on the singular geometries which create a new branch of edge states because of the inner lattice defect. The many-particle states of CI/QAH at the unit filling factor ($\nu=1$) can be obtained based on the Pauli principle. Charge fractionalization exists around the singular center and the fractional charge is revealed by the density profile along the radius. Interestingly, there are multiple branches of edge excitations in these systems because of more than one occupation choices for fermions meeting core states, and each branch of edge excitations has the same quasi-degeneracy sequence.

{\it Models and single-particle states.---}The kagom{\'e} lattice on disk geometry with $6$-fold rotational symmetry exhibits a nearly rounded edge [shown in Fig.~\ref{disk_Lattice}(a)]. However, when we reduce the rotational symmetry, the disk geometry changes into the conical surface and its unfolded drawing is a sector shown in Fig.~\ref{disk_Lattice}(b) with $5$-fold rotational symmetry. At the same time, when the rotational symmetry is increased, the disk geometry will be projected into a helicoid-like surface and its unfolded drawing is a disk and a sector with $7$-fold rotational symmetry shown in Fig.~\ref{disk_Lattice}(c). Here, we choose the sector with angle $\pi/3$ as a unit sector which is shown in Fig.~\ref{disk_Lattice}(c). Other singular geometries with arbitrary $n$-fold rotational symmetry can be constructed through ``cutting and gluing'' the unit sectors. In the following, we specify each singular lattice based on its radius and its $n$-fold rotational symmetry.

The tight-binding Hamiltonian of the kagom\'{e}-lattice CI/QAH model is given by:
\begin{eqnarray}
H_{\rm KG}= &-&t\sum_{\langle\mathbf{r}\mathbf{r}^{ \prime}\rangle}
\left[a^{\dagger}_{\mathbf{r}^{ \prime}}a_{\mathbf{r}}\exp\left(i\phi_{\mathbf{r}^{ \prime}\mathbf{r}}\right)+\textrm{H.c.}\right]\nonumber\\
&-&t^{\prime}\sum_{\langle\langle\mathbf{r}\mathbf{r}^{\prime}\rangle\rangle}
\left[a^{\dagger}_{\mathbf{r}^{\prime}}a_{\mathbf{r}}+\textrm{H.c.}\right]
\label{e.2}
\end{eqnarray}
Where $a^{\dagger}_{\mathbf{r}}$ ($a_{\mathbf{r}}$) creates (annihilates) a particle at site $\mathbf{r}$,$\langle\dots\rangle$ and $\langle\langle\dots\rangle\rangle$ denote the nearest-neighbor (NN) and the next-nearest-neighbor (NNN) pairs of sites. $\phi$ is the parameter for the staggered-flux phases. We choose the flat band parameters of the kagom\'{e} model: $t=1$, $t^{\prime}=-0.19$, $\phi=0.22\pi$~\cite{WWLuo,ALHe}. In order to obtain stable and confined edge states, we add the conventional harmonic trap on these geometries of the form $V = V_{\rm trap}\sum_{\mathbf{r}} |\mathbf{r}|^2 n_{\mathbf{r}}$ with $V_{\rm trap}$ as the potential strength (with the NN hopping $t$ as the energy unit), $|{\mathbf{r}}|$ as the radius from the disk center
(with the half lattice constant $a/2$ as the length unit)~\cite{WWLuo,ALHe}.

\begin{figure}[!htb]
  \vspace{-0.1in}
\includegraphics[scale=0.35]{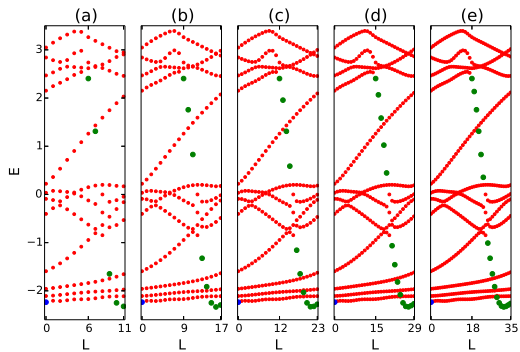}
  \vspace{-0.1in}
  \caption{(color online). (a)-(e) The single-particle energy spectra of the kagom{\'e} lattice on the singular geometries with $12$-, $18$-, $24$-, $30$- and $36$-fold rotational symmetry. There is a fixed-energy state in the $L=0$ angular momentum sector (blue dot) and some core states (green dot) around the corner in the gap and bulk for $12$-, $18$-, $24$-, $30$- and $36$-fold rotational symmetries. The green dot can be viewed as the ``edge state'' around the inner $n$-sided polygon. Here, the lattices on singular surfaces have the radius $r=5$ and the trap potential is $V_{\rm trap}=0.005$.}
\label{Energy_of_different_RS}
\end{figure}

The single-particle energy spectra and wave functions (or densities) of the Kagom{\'e}-lattice model can be obtained on the planar and singular surfaces. According to the results of numerical exact diagonalization, there are some core states on singular surfaces (shown in Fig.~\ref{single_particle_states}). Because of the increasing or reducing the number of unit sectors, lattice defects emerge around the lattice center with an edge formed around the inner $n$-sided polygon. These core states are confirmed by the single-particle density distributions in lattice sites. From the single-particle densities along the radial direction, it is possible to discern some states localized around the center. From Fig.~\ref{Energy_of_different_RS}, we observe that the number of core states is increased with the increasing number of unit sectors. Here, we choose the radius $r=5$ lattice with the trap potential $V_{\rm trap}=0.005$.  These core states are also the inner ``edge states'' around the lattice center because of the increasing the number of unit sectors, and the number of core states is related to the number $n$ for the system with the $n$-fold rotational symmetry.

\begin{figure}[!htb]
  \vspace{-0.1in}
\includegraphics[scale=0.55]{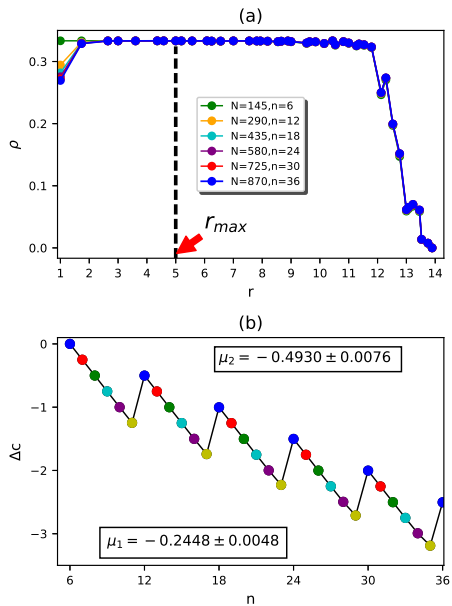}
  \vspace{-0.2in}
  \caption{(color online). (a) The density profile of the singular kagom{\'e} lattices along the radial direction. (b) The missing charge ($\Delta c$) varies periodically with the increasing of the number of unit sectors $n$. We use the linear function to fit their relations $\Delta c=\mu n+\beta$. Here, $\mu_1$ and $\mu_2$ denote the average slope of $\Delta c$ with adding a unit sector or adding a disk. The lattice radius is $r=14$ in an infinite potential well filled by $N=[145n/6]$ spinless fermions.}
\label{Charge}
\end{figure}

{\it Many particles and fractional charge.---}The wave function for IQH state on disk geometry is
\begin{equation}\label{IQH}
\Psi_{\rm{IQH}}(\{z_i\})=\displaystyle\prod_{i<j}(z_i-z_j){\rm{exp}}({-\displaystyle\sum_i|z_i|^2/4l_B^2}),
\end{equation}
where $z=x+iy$ and $l_B$ is the magnetic length. The IQH state of spinless fermions can be constructed based on the single-particle states and the Pauli principle. The observables (like the energy, density) of many-particle states are the superposition of the observable for single-particle states, $\mathcal{O} =\sum_{i} \mathcal{O}_{i}$, where $\mathcal{O}$ and $\mathcal{O}_i$ are the observables of many-particle and single-particle states (like the density profile shown in Fig.~\ref{Charge}(a)). For CI/QAH states, configurations of many-particle states are $111\ldots1,111\ldots101,111\ldots1001,\ldots$, where ``1'' and ``0'' denotes the particle number occupying the orbital. Many-particle states can be constructed based on the configurations.

\begin{figure*}
  \vspace{-0.1in}
\includegraphics[width=14cm]{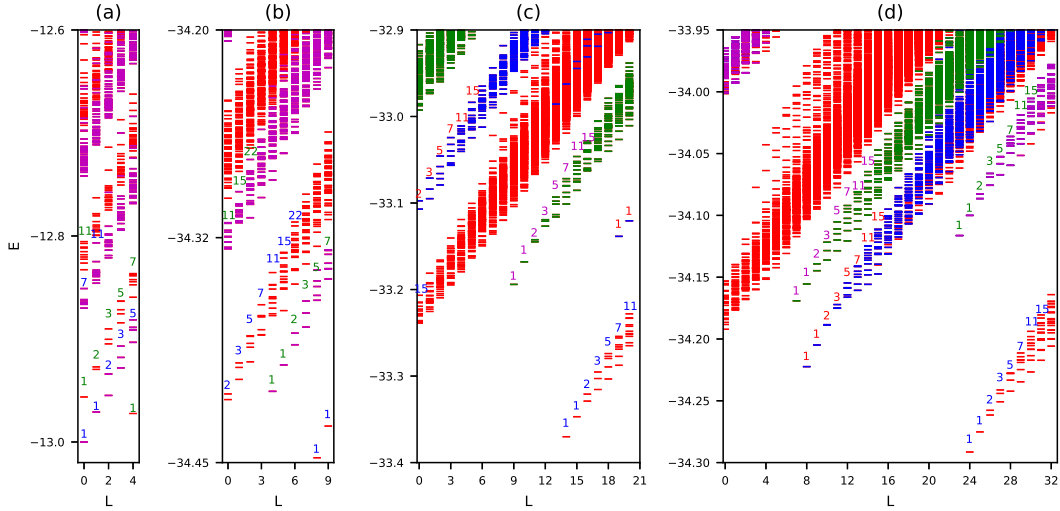}
  \vspace{-0.1in}
\caption{(color online). Edge excitations for the CI/QAH states on singular geometries with $5$-, $10$-, $21$- and $33$- fold rotational symmetry. (a)-(b) There are two branches of edge excitations for the CI/QAH with the fermions filling one core state. (c)-(d) There are four branches of edge excitations for the CI/QAH with the fermions filling two core states. There is an energy branch mixed with the red branch marked the blue color in (c). There number of filling spinless fermions is $6$ in (a) and $16$ in (b)-(d). Here, the lattice radius is $r=5$ with the trap potential $V_{\rm trap}=0.005$ in (a)-(b) and $V_{\rm trap}=0.015$ in (c)-(d).}
\label{Edge_Excitation}
\end{figure*}

Charge fractionalization has been observed on some non-fractional systems with defects like the SSH model~\cite{SSH}, the Kekul{\'e} lattice~\cite{Hou}, and the singular surfaces~\cite{Biswasa} and lattices~\cite{Ruegg,Moore}. There are some methods to calculate the fractional charge, e.g. from local density of states~\cite{Hou,Ruegg}, continuum description in the low energy limit~\cite{Ruegg,Moore}, and excessive charge in FQH states described in terms of the density profile~\cite{TuHH}.  Inspired by the calculation of excessive charge in FQH states on the singular surfaces~\cite{TuHH}, the fractional charge around the center of singular lattices is
\begin{equation}\label{FCH}
\Delta c=\sum_r^{r_{\rm max}}[\rho_r(n)-\rho_r(6)]n_r,
\end{equation}
where $r_{\rm max}$ is upper limit of summation, $\rho_r(n)$ is the density at the radial direction $r$ on the singular lattice with $n$-fold rotational symmetry and $n_r$ is the number of lattice sites at the radius $r$ [shown in Fig.~\ref{Charge}(a)].  Here, we choose $r=14$ singular lattices filling with $[145n/6]$ spinless fermions and `$[\dots]$' is the integer function. The fractional charge is fitted using a linear function which the slope $\mu_1$ is the missing charge when adds a unite sector and $\mu_2$ denotes the missing charge with adding a disk. Based on the fitting results, about $-1/4$, $-1/2$, $-3/4$, $\ldots$ fractional charge emerges around the center of the singular lattice [shown in Fig.~\ref{Charge}(b)] with adding a unit sector. For $n>6$ and $n<12$, the fractional charge shows $q_n=-(n-6)/4$  which is in accordance with the fractional charge in isolated defects on the lattice with conical singularities or topological fullerenes~\cite{Ruegg,Moore}. A reasonable explanation to the fractional charge is the excessive berry phase with the excessive area at the center of singular lattices~\cite{Moore}. Once adding a disk, a new core state (we call it ``effective core state'') will appear between the fixed-energy state (shown in Fig.~\ref{Energy_of_different_RS} blue dot) and the second energy band with a $\pm1/2$, $\pm1$, $\pm3/2$, ... charge emerging around the center of singular lattices.

{\it Edge excitations.---}Considering the configurations of many-particle states $111\ldots11,111\ldots101,111\ldots1001$, $\ldots$, a branch of edge excitations for fermions at integer filling on disk lattice can be obtained. Different from the CI/QAH states on disk, there are some core states in the band and gap for the CI/QAH states on singular surfaces (shown in Fig.~\ref{single_particle_states} and Fig.~\ref{Energy_of_different_RS}). The fermions occupy all energy levels obeying the Pauli principle including the core states at the case of integer filling. For a core state in the band, there are two cases for the fermions occupying the energy levels: occupying the core state or not. As a result, there are two branches of edge excitations when the fermions meeting a core state [shown in Fig.~\ref{Edge_Excitation}(a) and (b)]. Much more interestingly, when fermions meet two core states, there are four cases for the fermions occupying and four branches edge excitation [shown in Fig.~\ref{Edge_Excitation} (c) and (d)] can be observed on some singular lattices with the proper trap potential though some edge excitations may be mixed with others [like one shown in Fig.~\ref{Edge_Excitation}(c)]. Followed by analogy, ideally, there are $2^{N_{\rm core}}$ branches of edge excitations for a system with $N_{\rm core}$ core states in the energy band. These branches of edge excitations are different from some FQH states with two edge-excitation branches like in $2/5$ FQH states.  For these FQH states, two branches of edge excitations are observed because the FQH states consist of two independent droplets~\cite{Wen1,Wen2} instead of the core states.

It is surprising and interesting that the CI/QAH states
of integer filling have more than one branches on the singular lattices. For a $5$-fold rotational symmetry lattice with the trap potential $V_{\rm trap}=0.005$, the $7$th single-particle state is a core state with angular momentum $4$ and there are two branches of edge excitations in this system. One branch is the fermions only filling in the extended states [the purple branch in Fig.~\ref{Edge_Excitation}(a)] which the edge configurations are $111111\underline{\bf 0}$0, $111110\underline{\bf 0}1$, $111110\underline{\bf 0}01$, $111101\underline{\bf 0}1$, $111110\underline{\bf 0}001$, $1111010\underline{\bf 0}1$, $111011\underline{\bf 0}1$, etc. For the other branch (the red one), the fermion occupies in a core state with the energy configurations $111110\underline{\bf 1}0$, $111101\underline{\bf 1}0$, $111011\underline{\bf 1}0$, $111100\underline{\bf 1}1$, $111100\underline{\bf 1}010$, $111010\underline{\bf 1}10$, etc. Here, $\underline{\bf 1}$ ($\underline{\bf 0}$) denotes the fermions occupying (non-occupying) the core states. There are also two branches of edge excitations for some systems with other $n$-fold rotational symmetry on the singular lattices [shown in Fig.~\ref{Edge_Excitation}(b)]. Similarly, if a system exists two proper core states, the fermions have four choices to occupy with more than two branches of edge excitations. For example, there are two core states (the $7$th one and the 30th one) of the CI/QAH on the singular surface with $21$-fold rotational symmetry and the trap potential $V_{\rm trap}=0.015$. There are at least three obvious branches of edge excitations shown in Fig.~\ref{Edge_Excitation}(c) and the configurations for one of branches colored with red line are $111111\underline{\bf 1}1\ldots110$, $111111\underline{\bf 1}1\ldots1010$, $111111\underline{\bf 1}1\ldots10010$, $111111\underline{\bf 1}1\ldots0110$, etc. The edge configurations of fermions in the green branch without occupying the first core (the 7th) state are $111111\underline{\bf 0}1\ldots110$, $111111\underline{\bf 0}1\ldots1010$, $111111\underline{\bf 0}1\ldots10010$, $111111\underline{\bf 0}1\ldots0110$, etc. Some fermions occupy the second core (the $30$th) state and the energy configurations are $111111\underline{\bf 1}1\ldots1100\ldots0\underline{\bf 1}0$, $111111\underline{\bf 1}1\ldots10100\ldots0\underline{\bf 1}0$, $111111\underline{\bf 1}1\ldots10010\ldots0\underline{\bf 1}0$, $111111\underline{\bf 1}1\ldots10110\ldots0\underline{\bf 1}0$, etc. The $4$th edge excitations with the edge configuration $111111\underline{\bf 0}1\ldots10\ldots0\underline{\bf 1}0$, $\ldots$ are the blue ones covered in the red branch [shown in Fig.~\ref{Edge_Excitation}(c)]. Similarly, there are four obvious edge branches when fermions meet two proper core states [shown in Fig.~\ref{Edge_Excitation}(d)].

{\it Summary.---}The CI/QAH states on singular surfaces with arbitrary $n$-fold rotational symmetry can be obtained by varying the number of unit sectors. Different from the CI/QAH states on planar geometries, there are some core states localized around the center of the singular surfaces. We then construct the many-particle states and edge excitations at unit filling based on the Pauli principle. The charge fractionalization can be observed in the singular lattice at unit filling. For the cases with one core state, there are two branches of edge excitations. Fascinatingly, there are more than two branches of edge excitations observed when the number of core states increases.

{\it Acknowledgments.---}This work is supported by the NSFC of China Grants No.11374265 (Y.F.W.) and No.11274276 (C.D.G.), and the State Key Program for Basic Researches of China Grant No.2009CB929504 (C.D.G.).


\begin{thebibliography}{99}

\bibitem{Klitzing} K. v. Klitzing, G. Dorda and M. Pepper, Phys. Rev. Lett. {\bf 45}, 494 (1980).

\bibitem{Thouless} D. J. Thouless, M. Kohmoto, M. P. Nightingale, and M. den Nijs, Phys. Rev. Lett. {\bf 49}, 405 (1982).

\bibitem{Haldane} F. D. M. Haldane, Phys. Rev. Lett. {\bf 61}, 2015 (1988).



\bibitem{Yakovenko} V. M. Yakovenko,
Phys. Rev. Lett., {\bf 65}, 251 (1990). 
\bibitem{Nagaosa} K. Ohgushi, S. Murakami, and N. Nagaosa, Phys. Rev. B
{\bf 62}, R 6065 (2000). 
\bibitem{Lattice}X. L. Qi, Y. S. Wu, and S. C. Zhang, Phys. Rev. B {\bf 74}, 085308 (2006).
\bibitem{Kagome} E. Tang, J. W. Mei, and X. G. Wen, Phys. Rev. Lett. {\bf 106}, 236802 (2011);
T. Neupert, L. Santos, C. Chamon, and C. Mudry, Phys. Rev. Lett. {\bf 106}, 236804 (2011).
\bibitem{Checkerboard} K. Sun, Z. C. Gu, H. Katsura, and S. Das Sarma, Phys. Rev. Lett. {\bf 106}, 236803 (2011).

\bibitem{RealHaldane} G. Jotzu, M. Messer, R. Desbuquois, M. Lebrat,	T. Uehlinger, D. Greif	and T. Esslinger, Nature {\bf 515}, 237 (2014).


\bibitem{Tsui}
D. C. Tsui, H. L. Stormer and A. C. Gossard, Phys. Rev. Lett. {\bf 48} 1559 (1982).

\bibitem{Laughlin} R. B. Laughlin, Phys. Rev. Lett. {\bf 50}, 1395 (1983).

\bibitem{GPP}
 F. D. M. Haldane, Phys. Rev. Lett. {\bf 67} 937 (1991);
 B. A. Bernevig and F. D. M. Haldane, Phys. Rev. Lett. {\bf 100}, 246802 (2008); {\it ibid.} {\bf 101}, 246806 (2008);
 B. A. Bernevig and N. Regnault, Phys. Rev. Lett. {\bf 103},206801 (2009).

\bibitem{Milletari} M. Milletar{\`i} and B. Rosenow, Phys. Rev. Lett. {\bf 111}, 136807 (2013).
\bibitem{Inoue} H. Inoue, A. Grivnin, N. Ofek, I. Neder, M. Heiblum, V. Umansky, and D. Mahalu, Phys. Rev. Lett. {\bf 112}, 166801 (2014).
\bibitem{SSH} W. P. Su, J. R. Schrieffer and A. J. Heeger, Phys. Rev. Lett. {\bf 42}, 1698 (1979).
\bibitem{SSH1} W. P. Su, J. R. Schrieffer and A. J. Heeger, Phys. Rev. B {\bf 22}, 2099 (1980).
\bibitem{Hou} C. Y. Hou, C. Chamon and C. Mudry, Phys. Rev. Lett. {\bf 98}, 186809 (2007).


\bibitem{Biswasa} R. R. Biswas and D. T. Son, PANS {\bf 113}  8636 (2016).
\bibitem{Klevtsov1} S. Klevtsov, J. Phys. A: Math. Theor. {\bf 50} 234003 (2017).

\bibitem{Ruegg}A. R\"{u}egg and C. W. Lin, Phys. Rev. Lett. {\bf 110} 046401 (2013).
\bibitem{Moore}A. R\"{u}egg, S. Coh and J. E. Moore, Phys. Rev. B {\bf 88} 155127 (2013).


\bibitem{FQH_Sphere}  F. D. M. Haldane, Phys. Rev. Lett. {\bf 51}, 605 (1983).    
\bibitem{FQH_Cylinder}  D. J. Thouless, Surf. Sci. {\bf 142}, 147 (1984). 
\bibitem{FQH_Torus} F. D. M. Haldane and E. H. Rezayi, Phys. Rev. B {\bf 31}, 2529 (1985).  


\bibitem{TuHH} Y. H. Wu, H. H. Tu and G. J. Sreejith, Phys. Rev. A {\bf 96} 033622 (2017).
\bibitem{Klevtsov2} S. Klevtsov, arxiv:1608.02928.
\bibitem{TCan1}T. Can, M. Laskin, and P. Wiegmann, Phys. Rev. lett. {\bf 113} 046803 (2014).
\bibitem{TCan2} T. Can, Y. H. Chiu, M. Laskin and P. Wiegmann, Phys. Rev. lett. {\bf 117} 266803 (2016).
\bibitem{Gromv} A. Gromv, Phys. Rev. B {\bf 94} 085116 (2016).


\bibitem{FCI_Geometry} D. N. Sheng, Z. C. Gu, K. Sun, and L. Sheng,
Nature Commun. {\bf 2}, 389 (2011); 
Y. F. Wang, Z. C. Gu, C. D. Gong, and D. N. Sheng,
Phys. Rev. Lett. {\bf 107}, 146803 (2011); 
N. Regnault and B. A. Bernevig,
Phys. Rev. X {\bf 1}, 021014 (2011); 
Y. F. Wang, H. Yao, Z. C. Gu, C. D. Gong, and D. N. Sheng,
Phys. Rev. Lett. {\bf 108}, 126805 (2012); 
Y. L. Wu, N. Regnault, and B. A. Bernevig,
Phys. Rev. B {\bf 86}, 085129 (2012).

\bibitem{FCI_Geometry1}X. L. Qi, Phys. Rev. Lett. {\bf 107}, 126803 (2011);
C. H. Lee, R. Thomale, and X. L. Qi, Phys. Rev. B {\bf 88}, 035101 (2013).

\bibitem{WWLuo}W. W. Luo, W. C. Chen, Y. F. Wang and C. D. Gong,
Phys. Rev. B {\bf 88},161109(R)(2013).
\bibitem{ALHe} A. L. He, W. W. Luo, Y. F. Wang and C. D. Gong, New J. Phys.  {\bf 17} 125005 (2015).



\bibitem{Mobius1}L. T. Huang and D. H. Lee, Phys. Rev. B {\bf 84} 193106 (2011).
\bibitem{Mobius2}W. Beugeling, A. Quelle and C. Morais Smith,, Phys. Rev. B {\bf 89} 235112 (2014).


\bibitem{Halperin} B. I. Halperin, Phys. Rev. B {\bf 25}, 2185 (1982).



\bibitem{Wen1} X. G. Wen, Int. J. Mod. Phys. B {\bf 6}, 1711 (1992).
\bibitem{Wen2} X. G. Wen, Adv. Phys. {\bf 44}, 405 (1995).


\end{thebibliography}
\end{document}